\documentclass[sigconf, anonymous=false, nonacm=true]{acmart}
\AtBeginDocument{%
  \providecommand\BibTeX{{%
    \normalfont B\kern-0.5em{\scshape i\kern-0.25em b}\kern-0.8em\TeX}}}

\setcopyright{acmcopyright}
\copyrightyear{2021}
\acmYear{2021}
\acmDOI{}

\acmConference[ODD ’21]{ODD v6.0 at KDD 2021}{August 14--18, 2021}{Virtual Event, Singapore}
\acmBooktitle{Proceedings of ODD v6.0 Workshop on Outlier Detection De-constructed in Conjunction with ACM SIGKDD 2021, August 14--18, 2021, Virtual Event, Singapore (ODD v6.0 at KDD 2021)}
\acmPrice{15.00}
\acmISBN{}

\begin{document}

\title{Anomaly Detection and Automated Labeling for Voter Registration File Changes}


\author{Sam Royston}
\affiliation{%
  \institution{VoteShield, Protect Democracy}
  \city{New York}
  \country{USA}}
\email{sam.royston@voteshield.us}

\author{Ben Greenberg}
\affiliation{%
  \institution{VoteShield, Protect Democracy}
  \city{New York}
  \country{USA}}
\email{ben.greenberg@voteshield.us}

\author{Omeed Tavasoli}
\affiliation{%
  \institution{VoteShield, Protect Democracy}
  \city{New York}
  \country{USA}}
\email{omeed.tavasoli@voteshield.us}

\author{Courtenay Cotton}
\affiliation{%
  \institution{VoteShield, Protect Democracy}
  \city{New York}
  \country{USA}}
\email{courtenay.cotton@voteshield.us}

\renewcommand{\shortauthors}{Royston and Greenberg, et al.}

\begin{abstract}
Voter eligibility in United States elections is determined by a patchwork of state databases containing information about which citizens are eligible to vote.  Administrators at the state and local level are faced with the exceedingly difficult task of ensuring that each of their jurisdictions is properly managed, while also monitoring for improper modifications to the database. Monitoring changes to Voter Registration Files (VRFs) is crucial, given that a malicious actor wishing to disrupt the democratic process in the US would be well-advised to manipulate the contents of these files in order to achieve their goals. In 2020, we saw election officials perform admirably when faced with administering one of the most contentious elections in US history, but much work remains to secure and monitor the election systems Americans rely on.  Using data created by comparing snapshots taken of VRFs over time, we present a set of methods that make use of machine learning to ease the burden on analysts and administrators in protecting voter rolls.  We first evaluate the effectiveness of multiple unsupervised anomaly detection methods in detecting VRF modifications by modeling anomalous changes as sparse additive noise. In this setting we determine that statistical models comparing administrative districts within a short time span and non-negative matrix factorization are most effective for surfacing anomalous events for review. These methods were deployed during 2019-2020 in our organization's monitoring system and were used in collaboration with the office of the Iowa Secretary of State. Additionally, we propose a newly deployed model which uses historical and demographic metadata to label the likely root cause of database modifications. We hope to use this model to predict which modifications have known causes and therefore better identify potentially anomalous modifications.
\end{abstract}




\maketitle

\section{Introduction}
\label{sec:introduction}
Statewide Voter Registration Files (VRFs) contain information about every voter within a state and determine who is allowed to participate in elections.  As referenced in the U.S. Senate intelligence committee report released in Summer 2019, the infrastructure for managing these lists exposes a critical vulnerability in our electoral system \cite{senate_report}. By periodically taking snapshots of these files and analyzing the measured changes, we aim to automatically detect the nature of any modifications and flag those that require further review, adding important safeguards to the VRF maintenance process.

In this paper we present two main approaches to understanding the changes made to a voter file over time. In the first, we consider the (normalized) volume of different types of file changes as distributed over different locations and times. We apply several anomaly detection methods on these distributions in an unsupervised setting.  We apply basic statistical methods to detect outliers with respect to both time and location. We also 
apply low-rank models, exploiting correlations in the data in order to suggest less obvious anomalies.

In our second approach, we broaden our model to consider the characteristics of those voters whose records were modified. Using a combination of demographic and historical features, we describe groups of changed voters using their aggregate features. We believe these individual voter features should be informative about the type of change that happened to those voters and the reason for that change, assuming it was a valid maintenance action. We examine this assumption with labeled examples of several known causes of a particular change type (deactivation), and use these features to predict the labels. The ability to automatically distinguish among maintenance actions with known causes is another step forward in distinguishing the set of expected voter file changes from unexpected, potentially malicious changes.

Anomaly detection methods are difficult to evaluate since the problem domains are often missing concrete examples of anomalies to check against, and our case is no exception.  We only have access to a few anecdotal examples of erroneous deactivations and thankfully no examples of malicious modifications to the Voter File.  The space of hypothetical attacks to the Voter File is not easily modeled, for instance because an attack would likely be selected from a set of diverse strategies, might occur only once, and because the objectives of an attacker could range from simply disrupting an election, to affecting the outcome.  Further study, which is out of the scope of this work, is needed to model the many plausible types of attack; this is a main reason why in section \ref{sec:automatic_labeling} we focus on classifying the causes that we \textit{are} familiar with, rather than the ones we are not. 
While various methods have been applied to detecting suspicious election results in the political science literature, the focus of our work is specifically on monitoring changes to voter registration databases \textit{leading up to} an election. To our knowledge, the similar previous work has either not been applied in production, or has not focused specifically on using anomaly detection or supervised learning to help protect and maintain voter databases. Our contribution is a description of which anomaly detection methods work in this applied setting, providing baselines for future research, as well as a proposed automatic labeling system to help further reduce the burden on  administrators managing these databases.

\section{Prior Work}
\label{sec:prior_work}
In the field of political science, there has been some academic study of voter registration databases, but limited previous research into the dynamics of changes to these databases. Older works such as \cite{Stewart_2}, \cite{ansolabehere2010quality}, and \cite{alvarez2013evaluating} establish some baseline analyses of the data quality or lack thereof (in terms of availability, consistency, completeness, and accuracy) of voter registration data and voting records across states in the U.S. In \cite{merivaki2020our} Merivaki performs a case study in Missouri, modelling voter data error rates and suggesting a relationship between error rates and a county’s voter status data and other demographic data.

The Electronic Registration Information Center (ERIC) project \cite{becker2019innovation} is a non-partisan, non-profit organization which facilitates data sharing and analysis for its 29 member states. ERIC software matches duplicate voters on different states' VRFs so states can more effectively rid their voter files of stale records, in addition to a number of other useful administrative functions.  We find ERIC to be the most similar project to ours which is deployed and in active use, though their approach and functionality is based almost entirely on record linkage, rather than detection of anomalous changes. 

More recently, there have been some attempts to model how voter registration databases change over time and identify when changes are anomalous. In \cite{Stewart_1} Pettigrew and Stewart attempt to model and evaluate the accuracy of voter removal processes only, due to death or moving. In a case study of two states, they highlight the fact that different states have very different voter list maintenance procedures.

Later in \cite{Alvarez} Kim, Schneider, and Alvarez present a framework somewhat similar to the one we will present in section \ref{sec:temporal_anomalies}. They take many snapshots of a voter database over a period of time and identify the changes between them, in terms of records removed, added or changed. They suggest using an inter-quartile range (IQR) threshold applied to raw change counts across many snapshots in a single county in Florida to identify anomalies over time.  Furthermore, they confirm with the county’s registrar that the discovered anomalies in fact correspond to specific infrequent maintenance events. This contrasts with our work since one of our objectives is to identify anomalies that cannot be classified as routine maintenance, regardless of how infrequently some maintenances may occur.  In \cite{Alvarez_2} two of the same authors extend this work to modeling all counties in Florida over time. They derive parameters for a Bayesian model relating snapshot time, location (county), and county population size to voter record change rates. The authors then define anomalies as points with significant deviation from this model’s predictions. 

Anomaly detection models have also been applied in political science on election outcome data itself \cite{klimek2012statistical, zhang2019election}. More broadly, anomaly detection is a well-studied problem in many domains \cite{chandola2009anomaly}. The problem we study here bears some resemblance to fraud detection in financial domains \cite{ahmed2016survey} and database security problems \cite{kamra2008detecting}.
The techniques we use here range from straightforward statistical methods to matrix decomposition methods, which have often been applied to anomaly problems \cite{wang2004profiling, tonnelier2017anomaly, pascoal2012robust}. In section \ref{sec:automatic_labeling} we also present a classification task that we intend to use ultimately to aid in identifying anomalies, which also has precedent \cite{steinwart2005classification, ruff2019deep}, though not in the realm of VRF maintenance.

\section{Voter Registration Files}
\label{sec:vfprelim}
Voter Registration Files contain information about every voter who is registered in the state and sometimes voters who were registered in the past but have since been unregistered.  These files contain different information depending on the state, but generally all contain a set of fields describing each voter's name, address, age, political party, voter status, and history of participation in past elections.  We restrict our analyses to the voter file from the State of Iowa, though files from other states contain roughly the same information and there is no reason to believe that a variant of these methods would not be applicable in other states.  Only a subset of the fields describing each voter affect their ability to vote, including: \textbf{address fields}, \textbf{name fields}, \textbf{activation status},
\textbf{party affiliation} and of course their presence on the list in the first place (their \textbf{removal} or \textbf{registration} status).  We extract changes to these fields by computing differences between temporally adjacent versions of the file.  The change types that we use in our analyses are defined in table \ref{table:changetypes}. The modification data are inserted into a database and later aggregated by change date, locale (county), and change type for further analysis and anomaly detection, covered in sections \ref{sec:anomaly_detection} and \ref{sec:automatic_labeling}.

\begin{table}
\caption{Types of Voter Registration File changes which can affect someone's ability to vote}
\begin{tabular}{ |c|p{58mm}| } 
    \hline
    Change Type & Description \\
    \hline \hline
    Address & Changes to address fields, for example \textit{house-num, street-name, etc.} \\ \hline
    Name & Changes to \textit{first-name, middle-name, or last-name} fields. \\ \hline
    Removal &  A record is deleted (in the comparison no matching voter-id is found in the posterior file)  \\ \hline
    Registration & An insertion of a new record (no matching voter-id is found in the anterior file) \\ \hline
    Deactivation & A change in the \textit{voter-status} field from \textit{active to inactive}. Inactive voters can still vote, but are eligible for removal in maintenance processes. \\ \hline
    Activation & A change in the \textit{voter-status} field from \textit{inactive to active}  \\ \hline
        Party & A change in the \textit{party-affiliation} field \\ \hline
\end{tabular}
\label{table:changetypes}
\vspace{-0.3cm}
\end{table}

\section{Anomaly Detection}
\label{sec:anomaly_detection}
This section deals with analyzing change counts when measured in aggregate by locale. \footnote{A \textit{locale} is a geographic administrative unit, like a county}
Treating each (date, locale) pair as a single aggregate sample instead of dealing with individual records has a few benefits:

\begin{description}
    \item[Anonymity:] Input data do not contain any personally identifiable information.
    \item[Dataset size:] Aggregating by locale reduces computational load.
    \item[Actionable results:] There will be a responsible party for addressing any issues found.  
\end{description}

Moreover, administrative units are also the lens through which an attacker would view access to the VRF.  Different counties often use different database technologies than one another and have different security practices.  In the case of a hack, a likely target could be groupings of administrative units which have certain infrastructure configurations in common.  These sub-populations are assembled by aggregating modifications to the VRF by the date, administrative unit, and change type.  For example, all address changes in Polk County measured between January 3rd and January 10th, 2019 would constitute a sub-population of changes.

In the later sections \ref{sec:identify_event_labels} and \ref{sec:reallife}, we report these method's performance in detecting sparse artificial changes, as well as real-life examples of how they performed when applied to deactivations in the state of Iowa over a 2.5-year period spanning 2018 to 2020. In this section we describe the rationale of each of these methods in detail.

\subsection{The Modification Matrix} \label{sec:defn_mod_matrix}
\begin{figure}[htbp]
\centering
        % Place nodes
\hspace*{-1.2cm} 
\includegraphics[width=.63\textwidth]{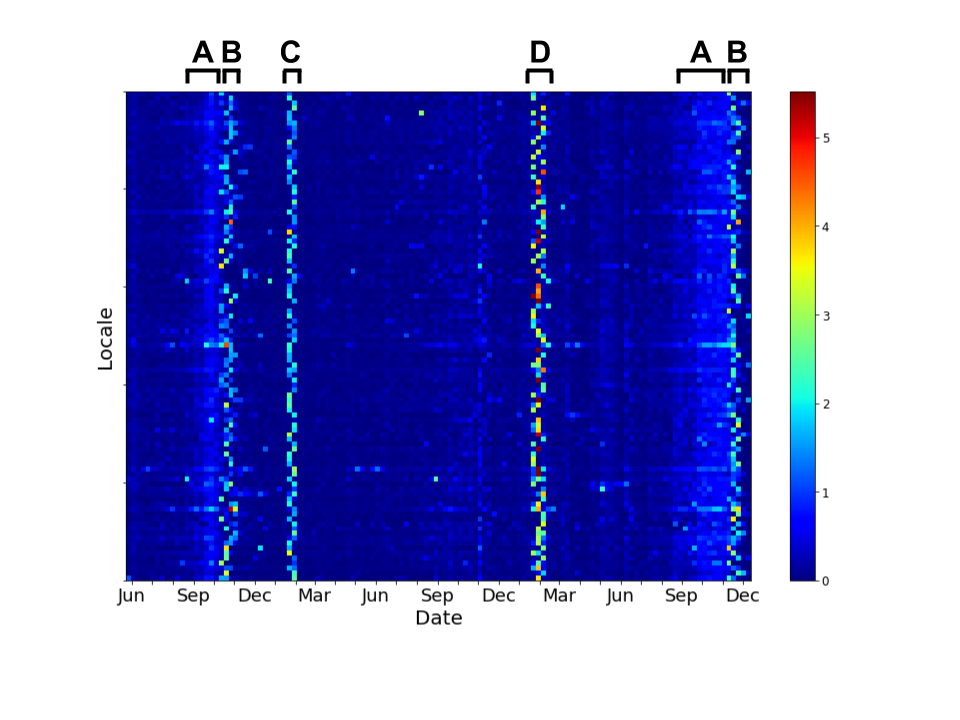}
\caption{The daily address change count per 1000 voters rendered as a modification matrix, where each row is a locale (i.e. county) and each column corresponds to a date. Some visible maintenance events are: 
       \textbf{(A)} Organic address changes via re-registration leading up to an election.
       \textbf{(B)} Same day address updates which took place on election day, later catalogued in the database.
       \textbf{(C)} Returned mailings are processed, resulting in address changes or deactivations
       \textbf{(D)} Addresses are updated after changes are recorded the day of the presidential primary, in addition to the return of the yearly list maintenance mailer
 }
\label{fig:mod_mtx}
\end{figure}

We will make frequent use of the \textit{modification matrix}, denoted $M$; a matrix of scalar values associated with each sub-population. All techniques discussed in this section operate on this matrix, whose entries are a normalized count of changes associated with each date and locale, for a given change type.  The normalization used is \textit{changes per-day, per one thousand voters}.  Consider the modification matrix $M \in \mathbb{R}^{\ell \times d}$ where $\ell$ is the number of locales and $d$ is the number of time intervals (rendered in figure \ref{fig:mod_mtx}). Each column is a different time
window over which modifications were tabulated and each row is a distinct locale.  We refer to administrative boundaries as locales, but any partitioning of the electorate can be used to populate the rows of M, and we suspect that alternative partitions may also be useful.

\subsection{Temporal Method}
\label{sec:temporal_anomalies}
This mode of anomaly detection stages the problem as one-dimensional time-series analysis.  Since we do
not have sufficient data to span multiple election cycles we must eschew seasonal models, and have no reason to believe that an
auto-correlative assumption is valid here. Based on this logic, we choose to evaluate simple univariate descriptive statistics based on
the Inter Quartile Range and standard deviation. 

\begin{equation}
    IQR(x) = \frac{x - Q_3}{Q_3 - Q_1} \;\;\;\;
    zscore(x) = \frac{x - \mu}{\sigma}
\end{equation}

where $\sigma$ is the standard deviation, $\mu$ is the arithmetic mean, and $Q_n$, $n \leq 4$ is the $n$th quartile. In temporal
anomaly detection we compute our descriptive statistics across each successive date range for each locale, e.g. there is a unique set of $Q$s
along with $\mu$ and $\sigma$ for each locale (and change type).
In our illustrations the row order is
alphabetical, but this ordering is never
relevant for the techniques we propose. For
temporal methods, ($\sigma, \mu, Q_n$) are
computed individually for each row, and does not change the order from if we just ranked the raw per-day, per-1000 voters score.  It's worth noting that the IQR method from \cite{Alvarez}
falls into this category.

In practice, the temporal method picks out county, date pairs primarily during the routine mailings of address confirmation postcards that took place in February of
2019 \textbf{(C)}.  This is a useful baseline which mimics what a human might do when inspecting a time series chart (one would pick out the spikes).  One would not want to rely exclusively on this method since it is predisposed to identify only large scale changes that are part of routine maintenance.

\subsection{Cross-Locale Method} 
\label{sec:cl_anomalies}

\begin{figure}
    \includegraphics[width=230px]{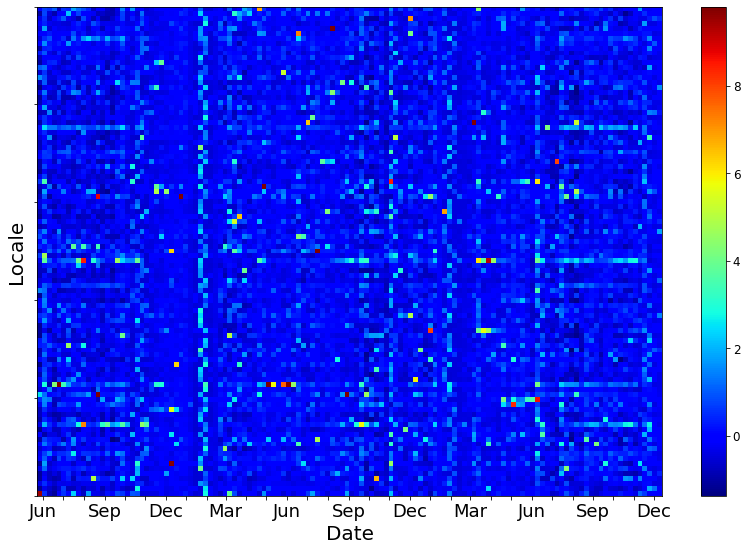}

    \caption{Cross-Locale anomaly scores applied to the modification matrix in figure \ref{fig:mod_mtx}, computed using the $STD$ method with a window of 5.}
\label{fig:cross_locale}
\end{figure}

When normalized by population, the maintenance behavior should be roughly identical (i.e. appear to draw from the same distribution) across all locales at a single point in time. Given this assumption, it makes sense to detect outliers by comparing behaviour across different locales within a restricted time range.  We use the same metrics listed in (1) but instead of computing  ($\sigma, \mu$ or $Q_n$) per row, we compute them over a sliding set of columns of width $2w + 1$ centered on the time interval corresponding to $x$.  The purpose of selecting $w > 0$ is to account for processes which may not co-occur exactly, but do co-occur within a limited amount of time. 

For $M_t \subset M, M_t \in \mathbb{R}^{\ell\times(2w + 1)}$ and with respect to $x$ we have, 

$$
M_t(x) = 
\begin{pmatrix}
    a_{1, col(x) - w} & \dots  & a_{1, col(x) + w} \\
    \vdots & \ddots & \vdots \\
    a_{\ell, col(x) - w} & \dots  & a_{\ell, col(x) + w}
\end{pmatrix} 
$$
Combining this definition with the IQR or Z-Score from (1), yields a complete description of the \textit{Cross-Locale} method. 
The IQRs and Z-Score\footnote{the terms Z-Score and STD are used interchangeably} methods yield similar results; the IQR method is superior when multiple anomalies co-occur temporally, while the Z-Score method is more sensitive to anomalies of lower magnitude. 
In our implementations we found that $w=2$ worked the best, which corresponds to sliding windows with a width of about 5 weeks.  See CL\_IQR\_5, CL\_IQR 3, CL\_STD\_5, and CL\_STD\_3, in figure \ref{fig:precision-gamma} and note that CL\_IQR\_$N$ corresponds to the full window width $N$ where $N = 2w + 1$. 

Qualitatively, this detection method is less activated when spikes happen en masse across counties (e.g. events \textbf{B} and \textbf{C}) and instead will highlight events that seem out of place compared to what else is going on at that point in time. Figure \ref{fig:cross_locale} shows an example of one of the Cross-Locale method's (STD with a window of 5) output anomaly scores.

\subsection{Low-Rank Models}
\label{sec:low_rank_models}
Modeling normal changes as a low rank matrix offers a more sophisticated approach to anomaly detection on $M$ than assuming each row or column represents an independent distribution of samples like we do in sections \ref{sec:temporal_anomalies} and \ref{sec:cl_anomalies}.

\begin{figure}
     \centering
    \includegraphics[width=230px] {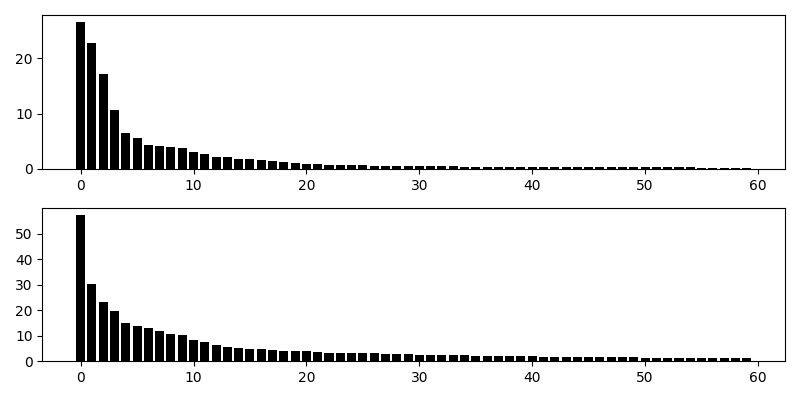}
    \caption{Top singular values of modification matrices for deactivations (top) and address changes (bottom).}
    \label{fig:svd}
\end{figure}

The top singular values seen in figure \ref{fig:svd} show us that these modification matrices are approximately low-rank (i.e. they have many singular values close to zero), so it makes sense to use models that rely on a low-rank assumption to model the different types of systematic maintenance.

\subsubsection{Non-Negative Matrix Factorization} \label{nmfanomalies}
\label{sec:nmf}

\begin{figure}
    \includegraphics[width=230px]{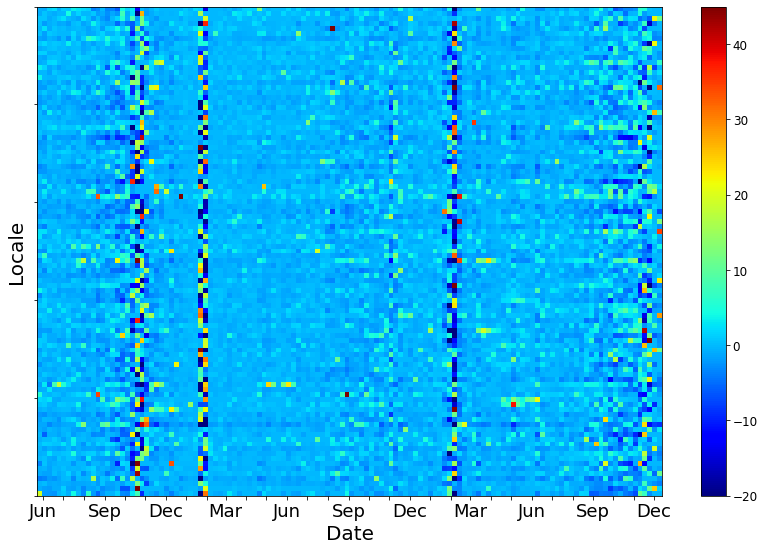}
    \caption{The matrix $M - PQ^\top$, from our NMF anomaly detection method, computed with $k = 5$.}
\label{fig:nmf}
\end{figure}

Non-negative matrix factorization \cite{nmf_3, nmf_1, nmf_2} involves finding a factorization of $M \in \mathbb{R}^{\ell \times d}$ consisting of two 
factor matrices $P \in \mathbb{R}^{\ell \times k}$ and $Q \in \mathbb{R}^{d 
\times k}$. The NMF anomaly detector assumes that the 
legitimate components of $M$ are the product of a limited number of factors associated with each date and locale and allows
explicit control over the number $k$ of latent factors.  Put another way, we can use $k$ to restrict the presumed rank of the matrix resulting from a legitimate process.  

This method also takes advantage of the non-negativity of $M$; since $M$ contains only normalized counts, all entries must be zero or greater.  This allows for restricting the factor matrices to be non-negative as well, resulting in fewer possible solutions to the optimization and (usually) better results \cite{nmf_1}.  $P, Q$ are selected to minimize the squared Frobenius norm (squared element-wise $\ell_2$ matrix norm) of the difference between $M$ and the factored reconstruction $PQ^\top$. 

We select $P, Q$ so as to
\begin{equation}
    \begin{aligned}
        &\text{minimize}\; \left\vert\left\vert M - PQ^\top \right\vert\right\vert_2^2 
        \\
        &\text{subject to }
        P > 0, \; Q > 0
    \end{aligned}
\end{equation}
and take the largest entries from $M - PQ^\top$ to be the most anomalous. Figure \ref{fig:nmf} shows as example of this difference output. We use the implementation in \cite{sklearn} which uses the coordinate descent method described in \cite{nmf_2} to solve (2).

\subsubsection{Robust Principal Component Analysis} \label{rpcaanomalies}
\label{sec:rpca}
Robust Principal Component Analysis \cite{RPCA} attempts to decompose an input matrix into the sum of two matrices, one low-rank and one sparse.  This formulation is uniquely appropriate since it seeks to model the underlying legitimate process and the anomaly process separately.  We perform anomaly detection by inspecting the highest-value entries of the sparse matrix.  Utilizing RPCA relies on applying a few generalizations about the structure of our modification matrix $M$. 
\begin{itemize}
    \item The matrix of \textit{natural} changes to the VRF is approximately low-rank.
    \item There are a small number of outliers within $M$ which do not adhere to this low-rank structure, i.e. the matrix of \textit{outliers} is sparse.
    \item The original matrix of changes is the sum of one \textit{natural} and one \textit{outlier} matrix $$ M = L + S $$
\end{itemize}
Where $S$ is sparse and $L$ is low-rank.  Finding such matrices is equivalent to solving the following optimization 
\begin{equation}
\begin{aligned} 
    \label{eq:rpca}
    \text{minimize}\;\;\;\; \left\vert\left\vert L \right\vert\right\vert_* + \lambda \left\vert\left\vert S \right\vert\right\vert_1 \\
    \text{subject to}\;\;\; M = L + S \;\;\;\;\;
\end{aligned}
\end{equation}

Where $\left\vert\left\vert L \right\vert\right\vert_*$ is the nuclear norm of $L$ \cite{sparsity}; a proxy for minimizing the rank directly (the latter is known to be computationally intractable), and $\left\vert\left\vert S \right\vert\right\vert_1$ is the entry-wise $\ell_1$ norm of $S$.  We used the implementation of the Iterative Shrinkage Thresholding Algorithm \cite{ISTA} found in \cite{ISTA_code} to solve equation \ref{eq:rpca} and perform anomaly detection by ranking the entries of $S$. 

\section{Automatic Labeling Using Population Metadata} \label{sec:automatic_labeling}

The techniques discussed in section \ref{sec:anomaly_detection} rely on structural properties of the modification matrices to highlight certain abnormal (date, locale, change type) groups in terms of a single metric: \textit{changes per-day, per one thousand voters}.  These techniques make no distinction between modifications which affect different demographic groups within a locale's population.

In this section, we present a framework for predicting anomalies using population metadata. Incorporating features such as age, electoral participation, and history of address changes has allowed us to observe statistical regularities at the population level, and identify as potential anomalies the (date, locale, change type) groups which differ from the model's expectation. The full set of features we use is listed in table \ref{table:variables}.

With access to VRF snapshots, we can use demographics, voting history, and change history data in order to predict anomalies. We aggregate voter data at the level of change groups; as in the modification matrix, our basic unit of population is the set of voters with modifications of some type, within some locale, between two consecutive snapshots. We compute the means of demographics- and history-derived features for the voters in each change group. These features may allow us to draw conclusions about the cause of the changes.

\subsection{Population Metadata}
\label{sec:population_metadata}

The availability of demographic variables varies by state -- the Iowa voter file tracks birthday, gender, voter status, and party identification. We treat age (in years) as a continuous numerical variable, and the others as one-hot encoded categorical variables. (E.g. for voter status, we use three features: \textit{voter status active}, \textit{voter status inactive}, and \textit{voter status pending}, with binary values representing the status of each voter).

We use features derived from individual voting histories, including numerical measures of turnout: participation (how often a voter turned out for elections in which they were eligible), partisanship (how often a voter turned out for their party), engagement (measured as the average size of elections a voter participated in), and provisional vote count. We also track temporal features, including days since last voted, months since registration, and months since the last VRF update.

Finally, the change history features measure how voters in a change group have been affected by each type of modification in the past. For each of the seven change types (see table \ref{table:changetypes}), we counted the number of changes affecting each voter during the prior 6 months (up to and not including the ``current'' change), as well as ``since the first VRF snapshot'' (again, not including the current change).

\begin{figure}
    \includegraphics[width=10cm]{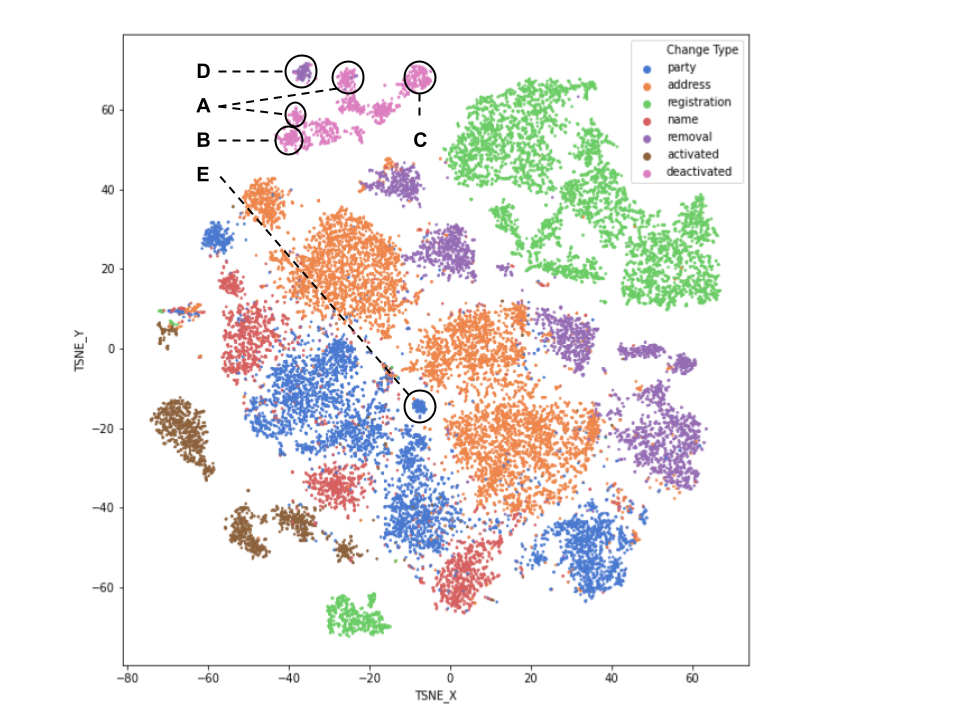}
    \centering
    \caption{Plot of a t-Stochastic Neighbor Embedding (t-SNE) \cite{tsne} of the features described in section \ref{sec:population_metadata}, colored by their change type. Labels are:
    (A) NCOA or Similar Mailing - Response Processing: 2019/2020 annual deactivations of non-respondents to election-related postcards 
    (B) Systematic September Maintenance 
    (C) Forwarded NCOA Mailing - No Response 
    (D) Inactivity Removals: Removal of voters due to inactivity (as opposed to death or incarceration) 
    (E) Libertarian Party Reclassification: Party changes caused by the reclassification of the Libertarian party to a political organization}
    \label{fig:tsne}
\end{figure}

\par
Concretely, for $n$ demographic features associated with each voter $\mathbf{v} \in \mathbb{R}^n$ we compute
$\mathbf{v_{\mu}} \in \mathbb{R}^n$ which is the column-wise mean of each voter within a single sub-population $p \in P$ where $P$ is the set of population partitions, defined by the date, locale, and change type.  Figure  \ref{fig:tsne} shows a t-Stochastic Neighbor Embedding created with this feature representation of (date, locale, change type) populations with a few interesting clusters highlighted, including the general region of some of the labels used by our supervised model (table \ref{table:event_labels}). The clear divisions between change types help validate the usefulness of these features.

\begin{comment}
When using the features listed in
table \ref{table:variables} we aim to make 
comparisons of the relative proportion of the general population 
being affected by the change, rather than the raw demographic 
proportions of the changed group. This is to say, we would like to 
allow our feature space the capacity to place change groups from 
administrative units with different demographic profiles within 
close proximity of one another when they are part of the same event class, such that the demographics of each change group are \textit{locale and date invariant}.

For every locale $\ell$ we have a vector of reference means measured from the population in that region $\mathbf{v}_{\ell} \in \mathbb{R}^n$ for that date, independent of any change; this is what we would expect if a group of changes were applied at random across a locale. A demographic feature vector is defined by the element-wise quotient $\mathbf{x} \in \mathbb{R}^n$, $\mathbf{x} \in X$, with $X \in \mathbb{R}^{\vert P\vert \times  n}$\footnote{Note that $\vert P\vert$ is the total number of entries in $M$.}
\begin{equation}
    \label{eq:quotient}
    \mathbf{x} = \frac{\mathbf{v_{\mu}}}{\mathbf{v_{\ell}}}
\end{equation}
\end{comment}

\begin{table}
\caption{Demographic and history features for Iowa voters}
\begin{tabular}{ |p{40mm}|c|c| } 
    \hline
    Variable & Type & Class \\
    \hline \hline
    months since registration & numerical & demographic \\ \hline
    years old & numerical & demographic \\ \hline
    gender & categorical & demographic \\ \hline
    voter status & categorical & demographic \\ \hline
    party membership & categorical & demographic \\ \hline
    days since last voted & numerical & voting history \\ \hline
    partisanship & numerical & voting history \\ \hline
    participation & numerical & voting history \\ \hline
    engagement & numerical & voting history \\ \hline
    provisional votes & numerical & voting history \\ \hline
    absentee votes & numerical & voting history \\ \hline
    <change type> changes last 6 months & numerical & change history \\ \hline
    <change type> changes all time & numerical & change history \\ \hline
\end{tabular}
\label{table:variables}
\end{table}

\subsection{Predicting Event Labels} \label{sec:predict_event_labels}
\par
Many modifications occur due to so-called ``systematic processes'' -- semi-regular VRF changes 
affecting a subset of voters in a locale. These processes define a pattern of normal behavior for the databases, and the corresponding modifications should not be flagged as anomalies. Therefore, we would like to predict the systematic process affecting a group of voters, using demographic and history data, in order to avoid false positives in anomaly detection. 
\par
Systematic processes are periodic VRF updates, initiated by state or local election officials in order to maintain an accurate voter file. For example, a state might periodically remove convicted felons from the voting roll as arrest records become available, or deactivate voters who fail to respond to a National Change of Address (NCOA) mailing. 
\par
For each change group, we calculated mean values of each demographic, voting history, and change history feature. After transforming to have zero-mean and unit-variance for the input, we trained an XGBoost classifier to predict which of the event labels would most likely explain the changes of the (locale, date) group in question.
\par
A robust label classifier can be used to complement an existing anomaly detection scheme. If the classifier predicts with high confidence that a set of modifications occurred due to a known pattern of behavior, an anomaly detection system should assign a lower probability that the modifications are anomalous. This creates a positive feedback loop; new user-provided event labels are used to train the classifier, which is in turn used to recommend examples in need of labeling.

\section{Experimental Evaluation}
\label{sec:groupeval}

\subsection{Improper Changes Modeled as Sparse Additive Noise} 
\label{sec:random_exp}
To our knowledge there exist few to no comparative evaluations of multiple anomaly detection techniques for VRDBs across the different types of changes which occur to VRFs, which we provide here.
\begin{table*}
\caption{AUC for each change type and strategy, for $\gamma$ ranging from 0 to 20x the average daily changes per 1000 voters for the relevant change type. 
This table is intended to facilitate comparisons between modification-ranking methodologies.}
\begin{tabular}{ c|p{8mm}|p{8mm}|p{8mm}|p{8mm}|p{8mm}|p{8mm}|p{18mm}|p{18mm}|p{13mm}|p{13mm} } 
        & NMF & RPCA & CL\newline STD 5 & CL\newline STD 3 & CL \newline IQR 5 & CL\newline IQR 3 & TEMPORAL \newline STD & TEMPORAL \newline IQR & GLOBAL \newline STD & GLOBAL \newline IQR \\ \hline
Deactivations & 0.38 & 0.23 & \textbf{0.72} & 0.66 & 0.03 & 0.02 & 0.03 & 0.04 & 0.02 & 0.02\\ \hline
Address & 0.59 & 0.41 & 0.74 & \textbf{0.76} & 0.33 & 0.38 & 0.18 & 0.18 & 0.17 & 0.19\\ \hline
Name & \textbf{0.66} & 0.41 & 0.63 & 0.66 & 0.3 & 0.31 & 0.57 & 0.31 & 0.28 & 0.27\\ \hline
Removal & \textbf{0.53} & 0.26 & 0.52 & 0.51 & 0.08 & 0.14 & 0.06 & 0.09 & 0.08 & 0.08\\ \hline
Activations & 0.59 & 0.4 & \textbf{0.63} & 0.6 & 0.37 & 0.37 & 0.43 & 0.32 & 0.33 & 0.33\\ \hline
Registrations & 0.74 & 0.57 & 0.83 & \textbf{0.84} & 0.45 & 0.49 & 0.43 & 0.38 & 0.39 & 0.38\\ \hline
Party & 0.66 & 0.39 & 0.82 & \textbf{0.84} & 0.36 & 0.38 & 0.06 & 0.06 & 0.05 & 0.07\\ \hline
\end{tabular}
\label{table:precision}
\end{table*}

In this section we evaluate our methods for detecting unexpected modifications in a modification matrix (described in section \ref{sec:anomaly_detection}). To do this, we perturbed $1\%$ of the entries at 
random in the modification matrix by adding a positive component $\gamma$.  For each change type, we tested each method's ability to
discern 
which entries were modified in $M$ by treating them as a ranking system: we took the top 20 highest ranking entries and reported the 
proportion of those which were members of the perturbed set.  As $\gamma$ becomes larger, 
the task becomes easier and easier, and for each change type we tested the methods over $\gamma$ ranging from $0$ to $20$ times the average daily changes per 1000 voters.  We based the $\gamma$ ranges on the averages to mitigate the effects of change frequency on the reported scores; otherwise for example, name changes would have appeared to be significantly easier to monitor than address changes because they are rarer.  Methods which achieve a higher precision at lower $\gamma$ are better at detecting this type of modification, and we use AUC to summarize each method's performance over the range of $\gamma$. 

It is important to recognise the major simplifications present in this evaluation. For one, we knew \textit{a priori} that there were greater than 20 perturbed entries, therefore the choice of selecting the top 20 suggested by the candidate methods is an artificially safe one.  In a real world use case we would not know much about the size of the perturbed set.  For this reason, these evaluations should be seen as a coarse affirmation of the ability to \textit{recommend} relevant anomalies for review, rather than an affirmation of our ability to detect anomalies.

Also included as baselines are global versions of the $STD$ and $IQR$ methods: GLOBAL STD and GLOBAL IQR, respectively.  These methods calculate the group statistics over the entire $M$, and since they both apply a single linear function to all values in M, they will both result in the same rankings as the raw input values.
In addition to these baselines, we included our methods based on {NMF} (Non-Negative Matrix Factorization, section \ref{nmfanomalies}), {RPCA} (Robust Principal Component Analysis, section \ref{rpcaanomalies}), {CL STD 5} (Cross-Locale Z-Score, with window width of 5 $w = 2$, section \ref{sec:cl_anomalies}), {CL STD 3} (Cross-Locale Z-Score, with window width of 3 $w = 1$, section \ref{sec:cl_anomalies}), {CL IQR 5} (Cross-Locale Inter-Quartile Range, with window width of 5 $w = 2$, section \ref{sec:cl_anomalies}), {CL IQR 3} (Cross-Locale IQR, with window width of 3 $w = 1$, section \ref{sec:cl_anomalies}), {TEMPORAL STD} (Temporal Z-Score, section \ref{sec:temporal_anomalies}), and {TEMPORAL IQR} (Temporal IQR, section \ref{sec:temporal_anomalies}).

Though the setup is slightly different since we have data from multiple counties, TEMPORAL IQR is essentially the same as the metric used in Kim, Schneider, and Alvarez \cite{Alvarez}.
Table \ref{table:precision} compares the AUC of the precision of each method as a function of $\gamma$ (defined above). In these experiments, Non-negative Matrix Factorization and the two Cross-Locale Z-Scores consistently outperform other methods across all $\gamma$. Refer to figure \ref{fig:precision-gamma} in the appendix for full plots. 

We find that cross-locale methods (specifically {CL STD 3} and {CL STD 5}, described in \ref{sec:cl_anomalies}) are successful in detecting anomalies among most change types, and are only outperformed in this context by NMF on detecting removals.  
% \begin{figure}
%     \centering
%     \includegraphics[width=230px]{results_address_anom.png}
%     \caption{Precision with which each anomaly model is able to identify perturbed entries (a random selection of $1\%$ of entries are increased by $\gamma$) in the address modification matrix within its top 20 suggested anomalies.}
%     \label{fig:random_noise_precision}
% \end{figure}

\subsection{Event Label Classification Results}
\label{sec:identify_event_labels}

\begin{table}
\caption{Event labels for systematic processes associated with deactivations}
\label{table:event_labels}
\begin{tabular}{ |p{23mm}|p{48mm}|p{8mm}| } 
    \hline
    Event Label & Description & Count \\
    \hline \hline
    Inactivity Mailing \newline - Response \newline Processing & A bulk deactivation due to processed responses (or lack thereof) from mailings sent by the government to ensure the veracity of address occupancy records & 99\\ \hline
    Systematic September Maintenance & An unspecified deactivation maintenance procedure which took place around Sep. 2019 & 37 \\ \hline
    NCOA Mailings & Similar to the inactivity mailing above, except that the mailings were sent as part of the maintenance of the National Change Of Address database & 27 \\ \hline
    Other & Any deactivation not included in the above classes, often composed of only a small number of changes & 21 \\ \hline
\end{tabular}
\end{table}

We trained a classifier to predict event labels for the voter features described in section \ref{sec:population_metadata} computed for deactivations in Iowa from June 2018 to Dec 2020. Deactivations are a step in the process of removing voters from VRFs who have not voted recently or did not respond to validation mail sent by election administrators. We used this change type for our labeling experiment since it usually only happens in large groups for specific reasons (in contrast to \textit{name} changes for example, which usually occur organically at the request of the voter). We hand-labeled 184 examples by examining the conditions of each deactivation group and validating our assessment with the office of the Iowa Secretary of State. Table \ref{table:event_labels} describes the four labels we applied and their counts in the dataset.

\begin{table}
\caption{Classification accuracy and F1 score on the holdout set for the label classifier}
\begin{tabular}{ c|c } 
    & Classification Performance \\ \hline
    Accuracy & 0.892 \\ \hline
    F1-Score & 0.892 \\ \hline
\end{tabular}
\label{table:labelResults}
\end{table}

\begin{comment}
We used the following normalization scheme:
\begin{itemize}
    \item For the categorical demographic, continuous demographic, and turnout features, we divided the raw change group features by the reference features. These values are centered around 1; e.g. a normalized feature value of 1.5 for the \textit{age\_years} feature indicates that the voters in this set of deactivated voters are on average 1.5 times as old as the registered voters in that locale, on that date.
    \item For the temporal features like \textit{days\_since\_last\_voted} and \textit{months\_since\_effective\_registration}, we chose to subtract the reference means (TODO: why?).
    \item For the change history features, we chose not to scale the change group features by the reference means; the base rate of modifications is so low (on the order of $10^{-3}$) that division by the reference mean produces high variance in the scaled features.
\end{itemize} 
\end{comment}

For each of the 184 groups of modified voters, we computed the demographics, voting history, and change history features using historical data. 20\% of the groups were reserved for the holdout set; the rest were used as inputs for training an XGBoost model with 50 estimators of depth 3, with a learning rate of 0.3.

Of the four labels, one (\textit{NCOA or Similar Mailing - Response Processing}) accounted for 99 of the 184 labels. We achieved a classification accuracy of 0.892, with an F1 score of 0.892, as seen in table \ref{table:labelResults}. Table \ref{table:confusion_matrix} shows the full confusion matrix from this experiment. At this level of accuracy, we are able to correctly identify the dominant event type affecting most groups of deactivations. As we only collected 184 labeled groups for this experiment, we hope that a larger dataset covering more locales, dates, and event types will make this classifier more effective for anomaly detection. Furthermore, we expect that predicting labels from modifications at the level of \textit{individual records}, as opposed to population groups, will yield a more finely-grained event label classifier.

\subsection{Identifying Real-Life Improper Deactivations}
\label{sec:reallife}

\begin{figure}
        \includegraphics[width=290px]{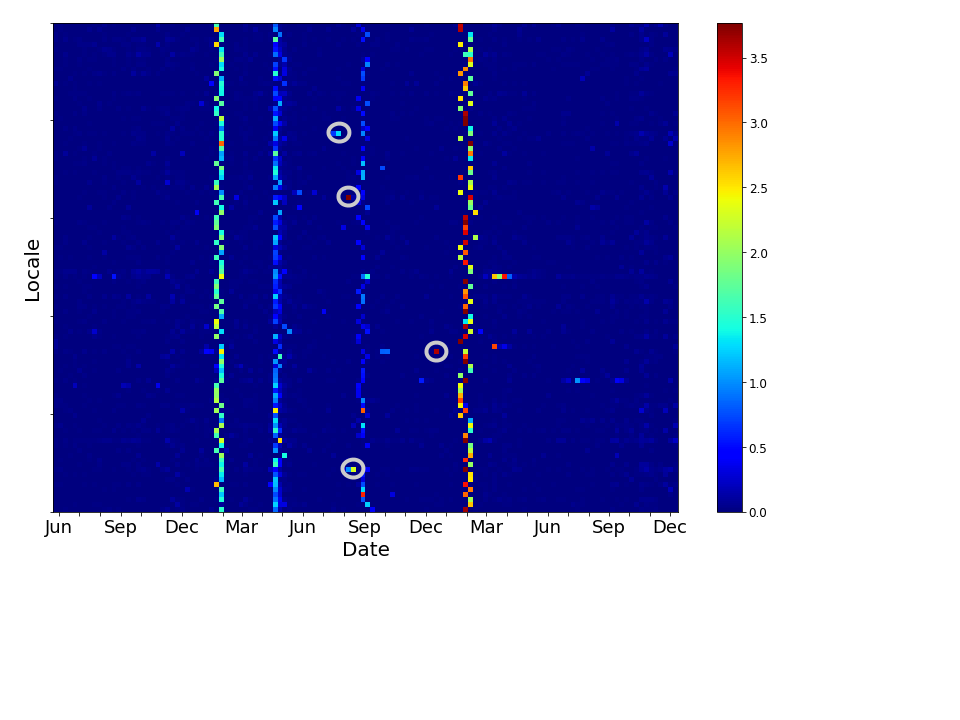}
        \vspace{-2cm}
        \caption{Due to a clerical error, four counties in Iowa mistakenly deactivated voters (the corresponding cells are outlined with white circles).  This error was corrected by the counties within a few weeks.}
        \label{fig:county_err}
\end{figure}

Due to a misunderstanding related to a school board election mailing, four counties in Iowa inadvertently deactivated several hundred voters in mid 2019. These deactivations are shown circled in figure \ref{fig:county_err}. These errors were detected by our platform and quickly corrected by the counties, but provide some anecdotal real-life performance data about our anomaly detection techniques.

To evaluate performance we took the scores assigned to the four sets of deactivations mentioned and ranked them relative to the scores for other dates and counties. We computed the average rank across the four examples for each anomaly detector; the best possible score (assuming these are the only four true anomalies) would have been 2.5 if the four examples occupy 1st, 2nd, 3rd, and 4th place, while the worst would be 14749.5 if the examples received the four lowest ranks.

\begin{figure}
        \includegraphics[width=200px]{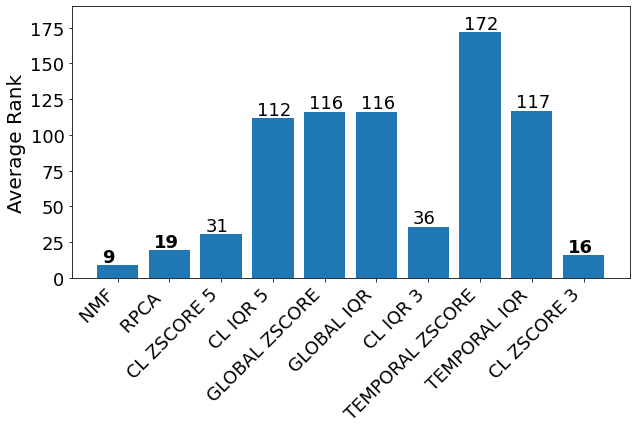}
        \caption{Average Ranking (out of 14751 entries) of the three improper deactivations shown in figure \ref{fig:county_err}, given by each anomaly detection technique (lower is better).}
        \label{fig:county_err_rank}
\end{figure}

The results can be found in figure \ref{fig:county_err_rank}, where NMF, Cross-Locale Z-Scores (width of 3), and RPCA all perform very well in ranking the improper deactivations as important relative to other population groups.

When features from these four change groups were fed into the event label classifier described in section \ref{sec:predict_event_labels}, they were predicted to have the label \textit{NCOA or Similar Mailing - Response Processing} with probabilities 0.995, 0.990, 0.987, and 0.952. These would be quickly labeled as anomalous, as NCOA mailings are generally processed during large, statewide maintenance events.

\section{Conclusion}
\label{sec:conclusion}
In this paper we present two broad models for understanding the dynamics of changes made to voter registration files. Both models begin with the same high level characterization of voter file changes as groups defined by their date, locale, and change type.

Our first type of model describes these change groups with a single normalized count feature. For each change type, these counts compose a modification matrix (of locales and dates) that can be operated on statistically or by using matrix decomposition methods to detect different types of unexpected values. We validated these models by adding random noise to certain matrix entries and evaluating the precision of each model in detecting those modified entries. Our NMF method and our Cross-Locale methods performed best on this synthetic noise task. Additionally, we have evidence from a small number of known real-life anomalous deactivation groups that our NMF, Cross-Locale Z-Scores, and RPCA methods would perform well in detecting those anomalies (and in fact did detect them in real life).

Our second class of model describes the same change groups with a more rich and descriptive feature set, taking demographic and history feature averages of the changed population. We evaluated this model on the more realistic and supervised task of classifying among four known reasons for voter deactivations, and were able to correctly predict these reasons with 89.2\% accuracy.

Both model types may be applied to the core problem of identifying anomalous changes in a VRF. The unsupervised models can be applied directly to rank change groups as potential anomalies, while the supervised model can help us learn common change patterns and thereby direct attention toward changes that do not fit any known pattern.  In the future, we plan to apply a similar labeling model to record level data such that we can predict the cause for each individual change to the VRF as opposed to causes associated with aggregate changes. We hope that our work inspires further interest in using machine learning to aid in the administration of Voter Registration Files. 

\begin{acks}
We would like to thank Carlos Fernandez-Granda of NYU for his valuable advice and feedback. We would also like to thank Wes Hicok and the Iowa Secretary of State’s Office, and Tim Hagle of the University of Iowa, for their help in understanding and labeling this data and for their feedback as early users of our anomaly detection system. Lastly, we would like to thank the entire VoteShield team for making this research possible.
\end{acks}

\bibliographystyle{ACM-Reference-Format}
\bibliography{main}


\begin{thebibliography}{28}

%%% the Web address.

\ifx \showCODEN    \undefined \def \showCODEN     #1{\unskip}     \fi
\ifx \showDOI      \undefined \def \showDOI       #1{#1}\fi
\ifx \showISBNx    \undefined \def \showISBNx     #1{\unskip}     \fi
\ifx \showISBNxiii \undefined \def \showISBNxiii  #1{\unskip}     \fi
\ifx \showISSN     \undefined \def \showISSN      #1{\unskip}     \fi
\ifx \showLCCN     \undefined \def \showLCCN      #1{\unskip}     \fi
\ifx \shownote     \undefined \def \shownote      #1{#1}          \fi
\ifx \showarticletitle \undefined \def \showarticletitle #1{#1}   \fi
\ifx \showURL      \undefined \def \showURL       {\relax}        \fi
% invisible to TeX
\providecommand\bibfield[2]{#2}
\providecommand\bibinfo[2]{#2}
\providecommand\natexlab[1]{#1}
\providecommand\showeprint[2][]{arXiv:#2}

\bibitem[\protect\citeauthoryear{Ahmed, Mahmood, and Islam}{Ahmed
  et~al\mbox{.}}{2016}]%
        {ahmed2016survey}
\bibfield{author}{\bibinfo{person}{Mohiuddin Ahmed},
  \bibinfo{person}{Abdun~Naser Mahmood}, {and} \bibinfo{person}{Md~Rafiqul
  Islam}.} \bibinfo{year}{2016}\natexlab{}.
\newblock \showarticletitle{A survey of anomaly detection techniques in
  financial domain}.
\newblock \bibinfo{journal}{\emph{Future Generation Computer Systems}}
  \bibinfo{volume}{55} (\bibinfo{year}{2016}), \bibinfo{pages}{278--288}.
\newblock


\bibitem[\protect\citeauthoryear{Alvarez, Ansolabehere, and
  Stewart~III}{Alvarez et~al\mbox{.}}{2004}]%
        {Stewart_2}
\bibfield{author}{\bibinfo{person}{R~Michael Alvarez}, \bibinfo{person}{Stephen
  Ansolabehere}, {and} \bibinfo{person}{Charles~H Stewart~III}.}
  \bibinfo{year}{2004}\natexlab{}.
\newblock \showarticletitle{Studying elections: Data quality and pitfalls in
  measuring the effects of voting technologies}.
\newblock  (\bibinfo{year}{2004}).
\newblock


\bibitem[\protect\citeauthoryear{Alvarez, Atkeson, and Hall}{Alvarez
  et~al\mbox{.}}{2013}]%
        {alvarez2013evaluating}
\bibfield{author}{\bibinfo{person}{R~Michael Alvarez},
  \bibinfo{person}{Lonna~Rae Atkeson}, {and} \bibinfo{person}{Thad~E Hall}.}
  \bibinfo{year}{2013}\natexlab{}.
\newblock \bibinfo{booktitle}{\emph{Evaluating elections: A handbook of methods
  and standards}}.
\newblock \bibinfo{publisher}{Cambridge University Press}.
\newblock


\bibitem[\protect\citeauthoryear{Amir~Beck}{Amir~Beck}{[n.d.]}]%
        {ISTA}
\bibfield{author}{\bibinfo{person}{Marc~Teboulle Amir~Beck}.}
  \bibinfo{year}{[n.d.]}\natexlab{}.
\newblock \showarticletitle{A Fast Iterative Shrinkage-Thresholding Algorithm
  for Linear Inverse Problems}.
\newblock
  \bibinfo{howpublished}{\url{https://www.math.ucdavis.edu/~sqma/MAT258A_Files/FISTA.pdf}}.
\newblock \bibinfo{journal}{\emph{SIAM Journal of Imaging Sciences}}
  (\bibinfo{year}{[n.\,d.]}).
\newblock


\bibitem[\protect\citeauthoryear{Ansolabehere and Hersh}{Ansolabehere and
  Hersh}{2010}]%
        {ansolabehere2010quality}
\bibfield{author}{\bibinfo{person}{Stephen Ansolabehere} {and}
  \bibinfo{person}{Eitan Hersh}.} \bibinfo{year}{2010}\natexlab{}.
\newblock \showarticletitle{The quality of voter registration records: A
  state-by-state analysis}.
\newblock \bibinfo{journal}{\emph{Report, Caltech/MIT Voting Technology
  Project}} (\bibinfo{year}{2010}).
\newblock


\bibitem[\protect\citeauthoryear{Becker}{Becker}{2019}]%
        {becker2019innovation}
\bibfield{author}{\bibinfo{person}{David Becker}.}
  \bibinfo{year}{2019}\natexlab{}.
\newblock \showarticletitle{Innovation in Synthesizing Big Data: The Electronic
  Registration Information Center (ERIC)}.
\newblock In \bibinfo{booktitle}{\emph{The Future of Election Administration}}.
  \bibinfo{publisher}{Springer}, \bibinfo{pages}{253--263}.
\newblock


\bibitem[\protect\citeauthoryear{Cao, Kim, and Alvarez}{Cao
  et~al\mbox{.}}{2020}]%
        {Alvarez_2}
\bibfield{author}{\bibinfo{person}{Jian Cao}, \bibinfo{person}{Seo-young~Silvia
  Kim}, {and} \bibinfo{person}{R~Michael Alvarez}.}
  \bibinfo{year}{2020}\natexlab{}.
\newblock \showarticletitle{Heterogeneity in Voter List Maintenance Practices:
  A Study of Florida Counties}.
\newblock  (\bibinfo{year}{2020}).
\newblock


\bibitem[\protect\citeauthoryear{Chandola, Banerjee, and Kumar}{Chandola
  et~al\mbox{.}}{2009}]%
        {chandola2009anomaly}
\bibfield{author}{\bibinfo{person}{Varun Chandola}, \bibinfo{person}{Arindam
  Banerjee}, {and} \bibinfo{person}{Vipin Kumar}.}
  \bibinfo{year}{2009}\natexlab{}.
\newblock \showarticletitle{Anomaly detection: A survey}.
\newblock \bibinfo{journal}{\emph{ACM computing surveys (CSUR)}}
  \bibinfo{volume}{41}, \bibinfo{number}{3} (\bibinfo{year}{2009}),
  \bibinfo{pages}{1--58}.
\newblock


\bibitem[\protect\citeauthoryear{Cichocki and Anh-Huy.}{Cichocki and
  Anh-Huy.}{2009}]%
        {nmf_1}
\bibfield{author}{\bibinfo{person}{Andrzej Cichocki} {and}
  \bibinfo{person}{P.~H. A.~N. Anh-Huy.}} \bibinfo{year}{2009}\natexlab{}.
\newblock \showarticletitle{Fast local algorithms for large scale nonnegative
  matrix and tensor factorizations.}
\newblock \bibinfo{journal}{\emph{IEICE transactions on fundamentals of
  electronics, communications and computer sciences}}  \bibinfo{volume}{92.3}
  (\bibinfo{year}{2009}), \bibinfo{pages}{708--721}.
\newblock


\bibitem[\protect\citeauthoryear{Daniel D.~Lee}{Daniel D.~Lee}{1999}]%
        {nmf_3}
\bibfield{author}{\bibinfo{person}{H.~Sebastian~Seung Daniel D.~Lee}.}
  \bibinfo{year}{1999}\natexlab{}.
\newblock \showarticletitle{Learning the parts of objects by non-negative
  matrix factorization}.
\newblock  (\bibinfo{year}{1999}).
\newblock


\bibitem[\protect\citeauthoryear{Emmanuel J.~Candes and Wright}{Emmanuel
  J.~Candes and Wright}{2011}]%
        {RPCA}
\bibfield{author}{\bibinfo{person}{Yi~Ma Emmanuel J.~Candes, Xiaodong~Li} {and}
  \bibinfo{person}{John Wright}.} \bibinfo{year}{2011}\natexlab{}.
\newblock \showarticletitle{Robust Principal Component Analysis?}
\newblock \bibinfo{howpublished}{\url{https://arxiv.org/pdf/0912.3599.pdf}}.
\newblock \bibinfo{journal}{\emph{Arxiv}} (\bibinfo{year}{2011}).
\newblock


\bibitem[\protect\citeauthoryear{Fevotte}{Fevotte}{2011}]%
        {nmf_2}
\bibfield{author}{\bibinfo{person}{J. Fevotte, C. \&~Idier}.}
  \bibinfo{year}{2011}\natexlab{}.
\newblock \showarticletitle{Algorithms for nonnegative matrix factorization
  with the beta-divergence.}
\newblock \bibinfo{howpublished}{\url{https://arxiv.org/pdf/1010.1763.pdf}}.
\newblock \bibinfo{journal}{\emph{Neural Computation}}  \bibinfo{volume}{23(9)}
  (\bibinfo{year}{2011}).
\newblock


\bibitem[\protect\citeauthoryear{Ganguli}{Ganguli}{2015}]%
        {ISTA_code}
\bibfield{author}{\bibinfo{person}{Dileep Ganguli}.}
  \bibinfo{year}{2015}\natexlab{}.
\newblock \bibinfo{title}{{Robust-PCA}}.
\newblock \bibinfo{howpublished}{\url{https://github.com/dganguli/robust-pca}}.
\newblock


\bibitem[\protect\citeauthoryear{Kamra, Terzi, and Bertino}{Kamra
  et~al\mbox{.}}{2008}]%
        {kamra2008detecting}
\bibfield{author}{\bibinfo{person}{Ashish Kamra}, \bibinfo{person}{Evimaria
  Terzi}, {and} \bibinfo{person}{Elisa Bertino}.}
  \bibinfo{year}{2008}\natexlab{}.
\newblock \showarticletitle{Detecting anomalous access patterns in relational
  databases}.
\newblock \bibinfo{journal}{\emph{The VLDB Journal}} \bibinfo{volume}{17},
  \bibinfo{number}{5} (\bibinfo{year}{2008}), \bibinfo{pages}{1063--1077}.
\newblock


\bibitem[\protect\citeauthoryear{Kim, Schneider, and Alvarez}{Kim
  et~al\mbox{.}}{2020}]%
        {Alvarez}
\bibfield{author}{\bibinfo{person}{Seo-young~Silvia Kim},
  \bibinfo{person}{Spencer Schneider}, {and} \bibinfo{person}{R~Michael
  Alvarez}.} \bibinfo{year}{2020}\natexlab{}.
\newblock \showarticletitle{Evaluating the Quality of Changes in Voter
  Registration Databases: Part of Special Symposium on Election Sciences}.
\newblock \bibinfo{journal}{\emph{American Politics Research}}
  \bibinfo{volume}{48}, \bibinfo{number}{6} (\bibinfo{year}{2020}),
  \bibinfo{pages}{670--676}.
\newblock


\bibitem[\protect\citeauthoryear{Klimek, Yegorov, Hanel, and Thurner}{Klimek
  et~al\mbox{.}}{2012}]%
        {klimek2012statistical}
\bibfield{author}{\bibinfo{person}{Peter Klimek}, \bibinfo{person}{Yuri
  Yegorov}, \bibinfo{person}{Rudolf Hanel}, {and} \bibinfo{person}{Stefan
  Thurner}.} \bibinfo{year}{2012}\natexlab{}.
\newblock \showarticletitle{Statistical detection of systematic election
  irregularities}.
\newblock \bibinfo{journal}{\emph{Proceedings of the National Academy of
  Sciences}} \bibinfo{volume}{109}, \bibinfo{number}{41}
  (\bibinfo{year}{2012}), \bibinfo{pages}{16469--16473}.
\newblock


\bibitem[\protect\citeauthoryear{Laurens van~der Maaten}{Laurens van~der
  Maaten}{2008}]%
        {tsne}
\bibfield{author}{\bibinfo{person}{Geoffrey~Hinton Laurens van~der Maaten}.}
  \bibinfo{year}{2008}\natexlab{}.
\newblock \showarticletitle{Visualizing Data Using t-SNE}.
\newblock  (\bibinfo{year}{2008}), \bibinfo{pages}{2579--2605}.
\newblock


\bibitem[\protect\citeauthoryear{Merivaki}{Merivaki}{2020}]%
        {merivaki2020our}
\bibfield{author}{\bibinfo{person}{Thessalia Merivaki}.}
  \bibinfo{year}{2020}\natexlab{}.
\newblock \showarticletitle{“Our Voter Rolls Are Cleaner Than Yours”:
  Balancing Access and Integrity in Voter List Maintenance}.
\newblock \bibinfo{journal}{\emph{American Politics Research}}
  (\bibinfo{year}{2020}), \bibinfo{pages}{1532673X20906472}.
\newblock


\bibitem[\protect\citeauthoryear{Pascoal, De~Oliveira, Valadas, Filzmoser,
  Salvador, and Pacheco}{Pascoal et~al\mbox{.}}{2012}]%
        {pascoal2012robust}
\bibfield{author}{\bibinfo{person}{Cl{\'a}udia Pascoal},
  \bibinfo{person}{M~Rosario De~Oliveira}, \bibinfo{person}{Rui Valadas},
  \bibinfo{person}{Peter Filzmoser}, \bibinfo{person}{Paulo Salvador}, {and}
  \bibinfo{person}{Ant{\'o}nio Pacheco}.} \bibinfo{year}{2012}\natexlab{}.
\newblock \showarticletitle{Robust feature selection and robust PCA for
  internet traffic anomaly detection}. In \bibinfo{booktitle}{\emph{2012
  Proceedings Ieee Infocom}}. IEEE, \bibinfo{pages}{1755--1763}.
\newblock


\bibitem[\protect\citeauthoryear{Pedregosa, Varoquaux, Gramfort, Michel,
  Thirion, Grisel, Blondel, Prettenhofer, Weiss, Dubourg, Vanderplas, Passos,
  Cournapeau, Brucher, Perrot, and Duchesnay}{Pedregosa et~al\mbox{.}}{2011}]%
        {sklearn}
\bibfield{author}{\bibinfo{person}{F. Pedregosa}, \bibinfo{person}{G.
  Varoquaux}, \bibinfo{person}{A. Gramfort}, \bibinfo{person}{V. Michel},
  \bibinfo{person}{B. Thirion}, \bibinfo{person}{O. Grisel},
  \bibinfo{person}{M. Blondel}, \bibinfo{person}{P. Prettenhofer},
  \bibinfo{person}{R. Weiss}, \bibinfo{person}{V. Dubourg}, \bibinfo{person}{J.
  Vanderplas}, \bibinfo{person}{A. Passos}, \bibinfo{person}{D. Cournapeau},
  \bibinfo{person}{M. Brucher}, \bibinfo{person}{M. Perrot}, {and}
  \bibinfo{person}{E. Duchesnay}.} \bibinfo{year}{2011}\natexlab{}.
\newblock \showarticletitle{Scikit-learn: Machine Learning in {P}ython}.
\newblock \bibinfo{journal}{\emph{Journal of Machine Learning Research}}
  \bibinfo{volume}{12} (\bibinfo{year}{2011}), \bibinfo{pages}{2825--2830}.
\newblock


\bibitem[\protect\citeauthoryear{Pettigrew and Stewart~III}{Pettigrew and
  Stewart~III}{2017}]%
        {Stewart_1}
\bibfield{author}{\bibinfo{person}{Stephen Pettigrew} {and}
  \bibinfo{person}{Charles Stewart~III}.} \bibinfo{year}{2017}\natexlab{}.
\newblock \showarticletitle{Moved out, moved on: Assessing the effectiveness of
  voter registration list maintenance}.
\newblock  (\bibinfo{year}{2017}).
\newblock


\bibitem[\protect\citeauthoryear{Ruff, Vandermeulen, G{\"o}rnitz, Binder,
  M{\"u}ller, M{\"u}ller, and Kloft}{Ruff et~al\mbox{.}}{2019}]%
        {ruff2019deep}
\bibfield{author}{\bibinfo{person}{Lukas Ruff}, \bibinfo{person}{Robert~A
  Vandermeulen}, \bibinfo{person}{Nico G{\"o}rnitz}, \bibinfo{person}{Alexander
  Binder}, \bibinfo{person}{Emmanuel M{\"u}ller}, \bibinfo{person}{Klaus-Robert
  M{\"u}ller}, {and} \bibinfo{person}{Marius Kloft}.}
  \bibinfo{year}{2019}\natexlab{}.
\newblock \showarticletitle{Deep semi-supervised anomaly detection}.
\newblock \bibinfo{journal}{\emph{arXiv preprint arXiv:1906.02694}}
  (\bibinfo{year}{2019}).
\newblock


\bibitem[\protect\citeauthoryear{Select Committee~on Intelligence}{Select
  Committee~on Intelligence}{2019}]%
        {senate_report}
\bibfield{author}{\bibinfo{person}{United States~Senate Select Committee~on
  Intelligence}.} \bibinfo{year}{2019}\natexlab{}.
\newblock \bibinfo{title}{RUSSIAN Active Measures, Campaigns, AND INTERFERENCE
  IN THE 2016 \text{U.S.} ELECTION, Volume 1}.
\newblock
  \bibinfo{howpublished}{\url{https://www.intelligence.senate.gov/sites/default/files/documents/Report_Volume1.pdf}}.
\newblock


\bibitem[\protect\citeauthoryear{Steinwart, Hush, and Scovel}{Steinwart
  et~al\mbox{.}}{2005}]%
        {steinwart2005classification}
\bibfield{author}{\bibinfo{person}{Ingo Steinwart}, \bibinfo{person}{Don Hush},
  {and} \bibinfo{person}{Clint Scovel}.} \bibinfo{year}{2005}\natexlab{}.
\newblock \showarticletitle{A classification framework for anomaly detection}.
\newblock \bibinfo{journal}{\emph{Journal of Machine Learning Research}}
  \bibinfo{volume}{6}, \bibinfo{number}{Feb} (\bibinfo{year}{2005}),
  \bibinfo{pages}{211--232}.
\newblock


\bibitem[\protect\citeauthoryear{Tonnelier, Baskiotis, Guigue, and
  Gallinari}{Tonnelier et~al\mbox{.}}{2017}]%
        {tonnelier2017anomaly}
\bibfield{author}{\bibinfo{person}{Emeric Tonnelier}, \bibinfo{person}{Nicolas
  Baskiotis}, \bibinfo{person}{Vincent Guigue}, {and} \bibinfo{person}{Patrick
  Gallinari}.} \bibinfo{year}{2017}\natexlab{}.
\newblock \showarticletitle{Anomaly detection and characterization in smart
  card logs using NMF and Tweets.}. In \bibinfo{booktitle}{\emph{ESANN}}.
\newblock


\bibitem[\protect\citeauthoryear{Trevor~Hastie}{Trevor~Hastie}{[n.d.]}]%
        {sparsity}
\bibfield{author}{\bibinfo{person}{Martin~Wainwright Trevor~Hastie,
  Robert~Tibshirani}.} \bibinfo{year}{[n.d.]}\natexlab{}.
\newblock \bibinfo{booktitle}{\emph{Statistical Learning with Sparsity: The
  Lasso and Generalizations}}.
\newblock \bibinfo{publisher}{Chapman `I\&' Hall/CRC Monographs on Statistics
  and Applied Probability}. 174--177 pages.
\newblock


\bibitem[\protect\citeauthoryear{Wang, Guan, and Zhang}{Wang
  et~al\mbox{.}}{2004}]%
        {wang2004profiling}
\bibfield{author}{\bibinfo{person}{Wei Wang}, \bibinfo{person}{Xiaohong Guan},
  {and} \bibinfo{person}{Xiangliang Zhang}.} \bibinfo{year}{2004}\natexlab{}.
\newblock \showarticletitle{Profiling program and user behaviors for anomaly
  intrusion detection based on non-negative matrix factorization}. In
  \bibinfo{booktitle}{\emph{2004 43rd IEEE Conference on Decision and Control
  (CDC)(IEEE Cat. No. 04CH37601)}}, Vol.~\bibinfo{volume}{1}. IEEE,
  \bibinfo{pages}{99--104}.
\newblock


\bibitem[\protect\citeauthoryear{Zhang, Alvarez, and Levin}{Zhang
  et~al\mbox{.}}{2019}]%
        {zhang2019election}
\bibfield{author}{\bibinfo{person}{Mali Zhang}, \bibinfo{person}{R~Michael
  Alvarez}, {and} \bibinfo{person}{Ines Levin}.}
  \bibinfo{year}{2019}\natexlab{}.
\newblock \showarticletitle{Election forensics: Using machine learning and
  synthetic data for possible election anomaly detection}.
\newblock \bibinfo{journal}{\emph{PloS one}} \bibinfo{volume}{14},
  \bibinfo{number}{10} (\bibinfo{year}{2019}), \bibinfo{pages}{e0223950}.
\newblock


\end{thebibliography}

\appendix

\section{Supplementary Materials}

\subsection{Artificial Change Detection Precision}

\begin{figure*}
    \centering
    \includegraphics[width=16cm]{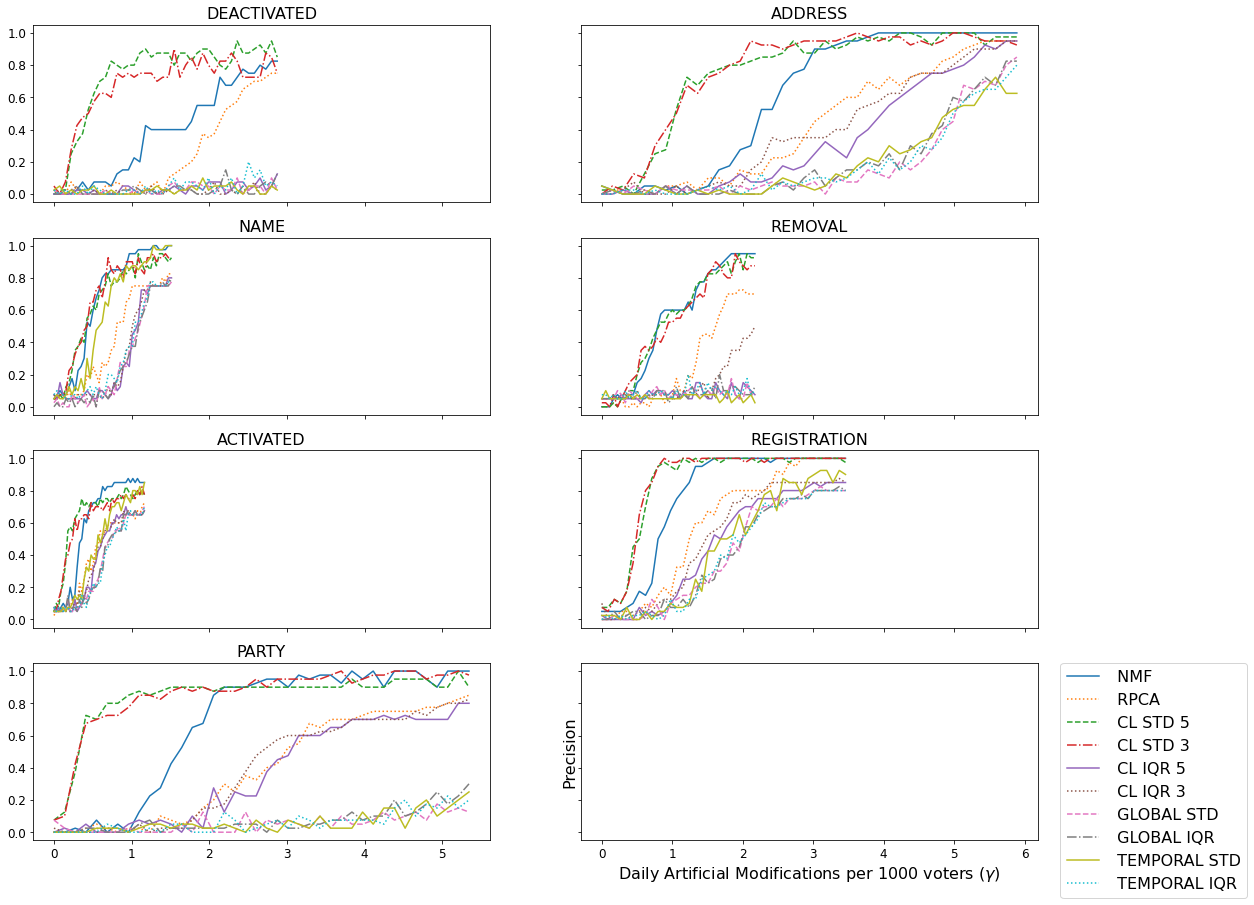}
    \caption{Precision in identifying artificial changes applied to $1\%$ of entries of the modification matrix, plotted as a function of $\gamma$, the magnitude of the artificial changes for each change type. The domain of each curve is always $[0, 20\mu_{M}]$ where $ \mu_{M}$ is the mean number of changes per day per 1000 voters for the relevant change type. Therefore, since name changes occur less frequently than address changes, the corresponding curves appear truncated.}
    \label{fig:precision-gamma}
\end{figure*}

See figure \ref{fig:precision-gamma}.

\subsection{Event Label Confusion Matrix}

See table \ref{table:confusion_matrix}.

\begin{table*}
\caption{Confusion matrix for event label classification. We saved 37 out of 184 labeled groups for the holdout set. Rows correspond to ground truth data; columns correspond to predicted labels.}
\begin{tabular}{ c|c|c|c|c } 
        & (1) & (2) & (3) & (4) \\ \hline
    (1) Response Processing & 18 & 0 & 1 & 0 \\ \hline
    (2) September & 1 & 8 & 0 & 0 \\ \hline
    (3) No Response & 0 & 0 & 6 & 1 \\ \hline
    (4) Other & 1 & 0 & 0 & 1 \\ \hline
\end{tabular}
\label{table:confusion_matrix}
\end{table*}

\end{document}